\documentclass[letterpaper,pra,twocolumn,showpacs,superscriptaddress]{revtex4}
\usepackage{amsmath}
\usepackage{amssymb}
\usepackage[dvips]{graphicx}
\usepackage[dvips,colorlinks=true,bookmarks=false,citecolor=blue,urlcolor=blue]{hyperref} %latex w/dvips

\begin{document}
%%%%%%%%%%%%%%%%%%%%%%%%%%%%%%%%%%%%%%%%%%%%%%%%%%%%%%%%%%%%%%%%%%%%%%%%%%%%%%%%%%%%%%
\title{Multi-Photon Quantum Key Distribution Based on Double-Lock Encryption}

\author{Kam Wai Clifford Chan}
\email{cliffchan@ou.edu}
\author{Mayssaa El Rifai}
\author{Pramode K. Verma}
\affiliation{\mbox{School of Electrical and Computer Engineering, University of Oklahoma -- Tulsa, Tulsa, OK 74135, USA}}
\author{Subhash Kak}
\affiliation{\mbox{School of Electrical and Computer Engineering, Oklahoma State University, Stillwater, OK 74078, USA}}
\author{Yuhua Chen}
\affiliation{\mbox{Department of Electrical and Computer Engineering, University of Houston, Houston, TX 77204, USA}}
\date{\today}


%%%%%%%%%%%%%%%%%%%%%%%%%%%%%%%%%%%%%%%%%%%%%%%%%%%%%%%%%%%%%%%%%%%%%%%%%%%%%%%%%%%%%%
\begin{abstract}
This paper presents a multi-stage, multi-photon quantum key distribution protocol based on the double-lock cryptography. It exploits the asymmetry in the detection strategies between the legitimate users and the eavesdropper. The security analysis of the protocol is presented with coherent states under the intercept-resend attack, the photon number splitting attack, and the man-in-the-middle attack. It is found that the mean photon number can be much larger than one. This complements the recent interest in multi-photon quantum communication protocols that require a pre-shared key between the legitimate users.
\end{abstract}

\maketitle
%%%%%%%%%%%%%%%%%%%%%%%%%%%%%%%%%%%%%%%%%%%%%%%%%%%%%%%%%%%%%%%%%%%%%%%%%%%%%%%%%%%%%%
\section{Introduction}\label{sec:intro}
%%%%%%%%%%%%%%%%%%%%%%%%%%%%%%%%%%%%%%%%%%%%%%%%%%%%%%%%%%%%%%%%%%%%%%%%%%%%%%%%%%%%%%
The security of quantum cryptography is based on the inherent uncertainty in quantum phenomena and it is the only known means of providing unconditionally secure communication~\cite{Gisin-etal2002, Scarani-etal2009}. Almost all contemporary practical QKD implementations are derived from the BB84 protocol~\cite{Bennett-Brassard1984}, a QKD protocol first proposed in 1984 and commercially implemented for limited market applications in the early 2000s. A major limitation of BB84 is that it requires no more than a single photon per time slot. Practically one relies on optical beams so weak that they generate less than one photon on average per time slot. In fact, BB84 and its decoy-state derivative are provably secure and optimal only when the mean photon number is about 0.5~\cite{Lo-Chau1999, Shor-Preskill2000, Gottesman-etal2004, Kraus-etal2007, Lo-etal2005, Ma-etal2005, Peng-etal2007}.

In this paper, we introduce a new multi-photon QKD protocol, first proposed in~\cite{Kak2006}, and demonstrate its security against the intercept-resend (IR) attack, the photon-number-splitting (PNS) attack, and the man-in-the-middle (MIM) attack. It is found that the mean photon number of the coherent pulses can generally be greater than 1. The protocol thus has the potential to allow QKD using detectors that may not be very efficient. It should be noted that multi-photon QKD protocols have also been studied under the category of continuous variable (CV) QKD protocols~\cite{Hillery2000, Cerf-etal2001, Grosshans-Grangier2002}. Nevertheless, CV-QKD requires homodyne or heterodyne detections, which are generally more complicated than photon counting used in discrete variable QKD protocols, as well as more sensitive to the noise in the quantum channel.

The principle behind the multi-photon, multi-stage QKD protocol to be discussed below is essentially the same as that of the classical double-lock cryptography. Security is given by the asymmetry in the detection strategies between the legitimate users and the eavesdropper, which is provided by the advantage creation akin to that utilized in the optimal quantum receiver in the Y00 (or $\alpha\eta$) protocol~\cite{Barbosa-etal2003} and the keyed communication in quantum noise (KCQ) method~\cite{Yuen2009}. A main difference is that the current protocol does not require a pre-shared key as in Y00 or KCQ.

In the following, we provide a detailed analysis on the security of the simplest form of the multi-stage protocol, the three-stage protocol. The dependence of the error probabilities in terms of the number of photons utilized in the channel will be studied.

%%%%%%%%%%%%%%%%%%%%%%%%%%%%%%%%%%%%%%%%%%%%%%%%%%%%%%%%%%%%%%%%%%%%%%%%%%%%%%%%%%%%%%
\section{Three-stage quantum cryptography}\label{sec:3stage}
%%%%%%%%%%%%%%%%%%%%%%%%%%%%%%%%%%%%%%%%%%%%%%%%%%%%%%%%%%%%%%%%%%%%%%%%%%%%%%%%%%%%%%
We now discuss the operations of the multi-photon tolerant quantum protocols in terms of transferring state $X$ from Alice to Bob. The state $X$ is one of two orthogonal states, designated as $|0\rangle$ and $|1\rangle$. The orthogonal states of $X$ represent 0 and 1 by prior mutual agreement of the parties and they are the cryptographic key being transmitted over the public channel. Alice and Bob apply secret transformations $U_A$ and $U_B$ which are commutative, i.e., $U_A U_B = U_B U_A$, on the quantum state before it is transmitted in the channel.

The steps of the protocol are described as follows:
\begin{enumerate}
\item
Alice applies a unitary transformation $U_A$ on information $X$ and sends it to Bob.

\item
Bob randomly chooses to retain the received signal $U_A(X)$ for authentication or to send it back to Alice with $U_B$ applied.

\item
Alice randomly chooses to retain the received signal $U_B U_A(X)$ for authentication or to send it back to Bob with $U_A^\dagger$ (complex conjugate transpose of $U_A$) applied.

\item
Bob applies $U_B^\dagger$ on $U_A^\dagger U_B U_A(X) = U_B(X)$ to get the information $X$.

\item
After receiving all the pulses, Bob announces publicly which pulses he has measured. Alice then discards those pulses that Bob did not measure. If the bit rate of the key is too low, the key is abandoned.

\item
Bob tells Alice the qubits he chose for authentication. Alice reveals to him the corresponding transformations and $X$ she applied for those qubits. The transformations are used to estimate the possibility of a man-in-the-middle attack. The authentication can likewise be performed using the pulses retained by Alice. Then Alice reveals to Bob some portion of the exchanged information $X$ to check the error rate.  They accept the rest of the key if the error rate of the key and the transformations are below certain thresholds.

\item
Alice and Bob finally perform post-processing (error correction and privacy amplification) as usual to minimize Eve's information.
\end{enumerate}

\begin{figure}[tb]
    \centering
    \includegraphics[height=4.7cm]{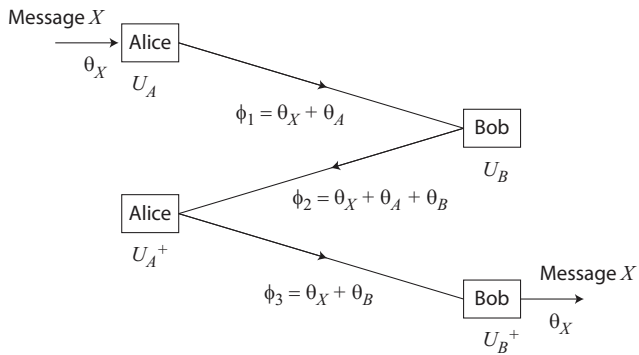}
    \caption{Schematic diagram of the three-stage protocol. The data $X$ is encrypted by the commutative transformations $U_A$ and $U_B$ during the transmission in the quantum channel.}
\label{fig:Figure1_3stage}
\end{figure}

The schematic of the three-stage protocol is shown in Fig.~\ref{fig:Figure1_3stage}. It should be noted that the unitaries $U_A$ and $U_B$ can be some general transformations that can mask the original state $X$. The parameters in $U_A$ and $U_B$ can be from a discrete or a continuous set. In the following, we consider the encoding of the qubit using the polarization of the photons in a coherent state. In this way, the data $X$ is represented by the polarization angle $\theta_X$ and the unitary transformations $U_A$ and $U_B$ are realized as polarization rotations through angles $\theta_A$ and $\theta_B$ respectively. Note that since $U_A$ and $U_B$ are required to be commutative, the polarization rotations are constrained on a great circle of the Poincar\'{e} sphere.

It is remarked that the use of many polarization angles to increase the robustness of QKD protocols has also been explored in~\cite{Acin-etal2004, Grazioso-Grosshans2013}. In addition, the scheme described here can in principle be applied to other degrees of freedom of the photons just like the Y00 protocol~\cite{Corndorf-etal2005, Harasawa-etal2011} and the state $X$ can be extended to multi-alphabet qudit systems.

%%%%%%%%%%%%%%%%%%%%%%%%%%%%%%%%%%%%%%%%%%%%%%%%%%%%%%%%%%%%%%%%%%%%%%%%%%%%%%%%%%%%%%
\section{Detailed Analysis of the Three-stage Protocol}\label{sec:analysis}
%%%%%%%%%%%%%%%%%%%%%%%%%%%%%%%%%%%%%%%%%%%%%%%%%%%%%%%%%%%%%%%%%%%%%%%%%%%%%%%%%%%%%%
The three-stage protocol can be viewed as a quantum double-lock encryption, with the unitary transformations (polarization rotations in this case) $U_A$ and $U_B$ acting as the locks. Therefore, the security of the three-stage protocol relies on (1) the ability of the transformations in protecting the transmitted bit value, and (2) the premise that Alice and Bob are certain that they are the ones who really applied the locks during the information transmission. The man-in-the-middle (MIM) attack by an intruder Eve exploits the second condition. We shall treat the two situations in detail in the following.

The main idea of the three-stage protocol is that, to obtain the key $X$, Bob only needs to distinguish between two possible orthogonal polarization states, e.g., horizontal or vertical. On the other hand, in order to obtain useful information, an eavesdropper Eve needs to determine the arbitrary unknown polarization angles $\phi_i$ of the quantum state transmitting in the quantum channel. The accuracy of her measurement depends on the number of photons she can access from the channel.

The amount of polarization rotations that both sides (Alice and Bob) select to give to the information bits is an arbitrary and independent value which varies from 0 to 180 degrees. It should be noted that practically one must also take into account the difficulties with maintaining stability and fidelity in the presence of noise. In the following analysis, we will ignore such requirements and assume that Alice and Bob can maintain perfect alignment in their basis for simplicity.

%%%%%%%%%%%%%%%%%%%%%%%%%%%%%%%%%%%%%%%%%%%%%%%%%%%%%%%%%%%%%%%%%%%%%%%%%%%%%%%%%%%%%%
\subsection{Intercept-Resend (IR) and Photon Number Splitting (PNS) Attacks}\label{sec:IR_PNS}
%%%%%%%%%%%%%%%%%%%%%%%%%%%%%%%%%%%%%%%%%%%%%%%%%%%%%%%%%%%%%%%%%%%%%%%%%%%%%%%%%%%%%%
We first consider the situation that the communication between Alice and Bob is “authenticated”, i.e., Alice knows that the information she sends out passes through Bob in the intermediate step and vice versa. Under this assumption, Eve can launch intercept-resend (IR) attacks, or more importantly, the photon-number-splitting (PNS) attack.

The main difference between IR and PNS is that, in IR all the photons are being taken out by Eve and she then resends any photon state to Bob. On the other hand, under PNS attack, the number of photons Bob receives is less than that in the original pulse. Such a loss of photons practically could be due to the channel loss, but in the analysis one has to attribute it to the action of the intruder.

If we further restrict ourselves to the situation of incoherent attacks, Eve is required to perform measurements before the classical postprocessing. Therefore, under the IR attack, the polarization states of the pulses resent by Eve usually are different from that she intercepts because of the measurement process, with the difference depending on the number of photons she receives. We can estimate the security against the IR attack by assuming the polarization states of the resent pulses are the same as those before the pulses are intercepted by Eve. This is an overestimation of the ability of Eve. Nevertheless, it enables us to analyze IR and PNS attacks using the same formalism. In addition, Eve's information $I_{EA}$ and $I_{EB}$ become identical, where $I_{EA} = \max_\text{Eve} I(E:A)$ is the maximal mutual information between Eve and Alice with a similar expression for $I(E:B)$.

For the IR and PNS attacks on the three-stage protocol, Eve only needs to measure the polarization angles of any two stages (say the first and the second stages). Then she can extract the bit value by orienting her measurement device in the third stage according the angles of the first and second stages.

More definitely, suppose the polarization angles of the three stages of the protocol are denoted by $\phi_1 = \theta_X + \theta_A$, $\phi_2 = \theta_X + \theta_A + \theta_B$ and $\phi_3 = \theta_x + \theta_B$, where $\theta_X$ is the information bit angle (0 or $\pi/2$), and $\theta_A$ and $\theta_B$ are the angles associated with Alice's and Bob's unitary transformations. Then the corresponding angles estimated by Eve for the first two stages are written as $\hat{\phi}_1 = \hat{\theta}_X + \hat{\theta}_A$ and $\hat{\phi}_2 = \hat{\theta}_X + \hat{\theta}_A + \hat{\theta}_B$. As a result, $\hat{\theta}_B = \hat{\phi}_2 - \hat{\phi}_1$. In order for Eve to obtain useful information, she requires that the error in determining $\theta_B$ should not be too large. Since $\theta_X$ is a binary random number of either 0 or $\pi/2$, Eve will determine the bit value erroneously if $\left|\hat{\theta}_B - \theta_B\right| = \left|\left(\hat{\phi}_2 - \hat{\phi}_1\right) - \left(\phi_2 - \phi_1\right)\right| > \pi/4$. The error probability of Eve is then given by
\begin{eqnarray}
    P_e(N_1, N_2)
    \hspace{-1mm}
    &=&
    \hspace{-1mm}
    \int_S d\hat{\phi}_1 \, d\hat{\phi}_2 \, d\phi_1 \, d\phi_2 \
    P(\phi_1) P_1(\hat{\phi}_1 | \phi_1, N_1)
\nonumber\\
&&
    \times\,
    P(\phi_2) P_2(\hat{\phi}_2 | \phi_2, N_2)
    ,
\label{eq:Pe_IR-PNS}
\end{eqnarray}
where $P(\phi) = 1/2\pi$ is the prior distribution of Alice's (Bob's) rotation angle and $P_i(\hat{\phi}_i | \phi_i, N_i)$ is the conditional probability of determining $\hat{\phi}_i$ given the angle $\phi_i$ and the mean photon number $N_i$ that is accessible by Eve. The integration domain $S$ corresponds to the region where the condition $\left|\left(\hat{\phi}_2 - \hat{\phi}_1\right) - \left(\phi_2 - \phi_1\right)\right| > \pi/4$ is satisfied. The mutual information $I(E:A)$ is given by $I(E:A) = 1 - h(P_e)$, where $h(x)$ is the binary entropy function.

Consider a three-stage protocol using coherent states of mean photon number $N$. First of all, Alice should randomize the phases of the coherent states to avoid Eve exploiting the phase information~\cite{Zhao-etal2007}. In this case, the quantum state is described by a density matrix with photon number following the Poisson distribution with parameter $N$.

To obtain a bound of the secure key rate, one has to estimate Eve's maximal information (see Section~\ref{sec:key_rate} below). This involves an optimal measurement strategy to obtain the conditional probability $P_i(\hat{\phi}_i | \phi_i, N_i)$. Bagan et al.~\cite{Bagan-etal2005} gave a detailed comparison of the estimation of the polarization state of a finite number of photons using the collective and local measurements. Instead of an optimal polarization measurement, in the following we consider a simple strategy that Eve performs polarization analysis with a fixed basis, denoted as horizontal and vertical, that is the same as Alice's and Bob's basis. Such a fixed basis (or tomographic) measurement is generally not optimal. Nevertheless, we additionally assume that Eve can determine the polarization angle correctly using a single basis only, instead of two bases that are required for the polarization states on a circle of the Poincar\'{e} sphere. This is accomplished by attributing Eve's measured polarization in the correct quadrant as the original polarization in the numerical calculations below. This procedure \emph{effectively doubles the number of photons available to Eve} for the estimation, and the fidelity obtained is generally even better than that using optimal collective measurements (see Fig. 1 of~\cite{Bagan-etal2005}).

With the measurement strategy mentioned above, the probability distributions of Eve's numbers of horizontal and vertical photons in the three stages are given by
\begin{eqnarray}
&&\hspace{-8mm}
    P_i(n_{H,i}, n_{V,i} | \phi_i, N_i)
\nonumber\\
    \hspace{-1mm}
    &=&
    \hspace{-1mm}
    \frac{e^{-N_i}}{1-e^{-N_i}}
    \frac{\left(N_i \cos^2\phi_i\right)^{2n_{H,i}} \left(N_i \sin^2\phi_i\right)^{2n_{V,i}}}
    {n_{H,i}! n_{V,i}!}
    ,
\label{eq:P_conditional_nHnV}
\end{eqnarray}
for $i = 1, 2, 3$, where $N_i$ is the mean number of photons in stage $i$ that is accessible by Eve. Here the continuous variable $\hat{\phi}_i$ in Eq.~(\ref{eq:Pe_IR-PNS}) is replaced by the discrete variables $n_{H,i}$ and $n_{V,i}$. Then $\phi_i$ can be estimated from the numbers of photons detected in the vertical port ($n_{V,i}$) and the horizontal port ($n_{H,i}$) of the polarization analyzer by $\tan^2\hat{\phi}_i = n_{V,i} / n_{H,i}$. Note that in Eq.~(\ref{eq:P_conditional_nHnV}), $n_{H,i}$ and $n_{V,i}$ cannot be zero simultaneously, for this gives no information to Eve about the angle $\phi_i$. Also we assume $N_i$ is known to Eve.

\begin{figure*}[!t]
    \centering
    \includegraphics[height=6cm]{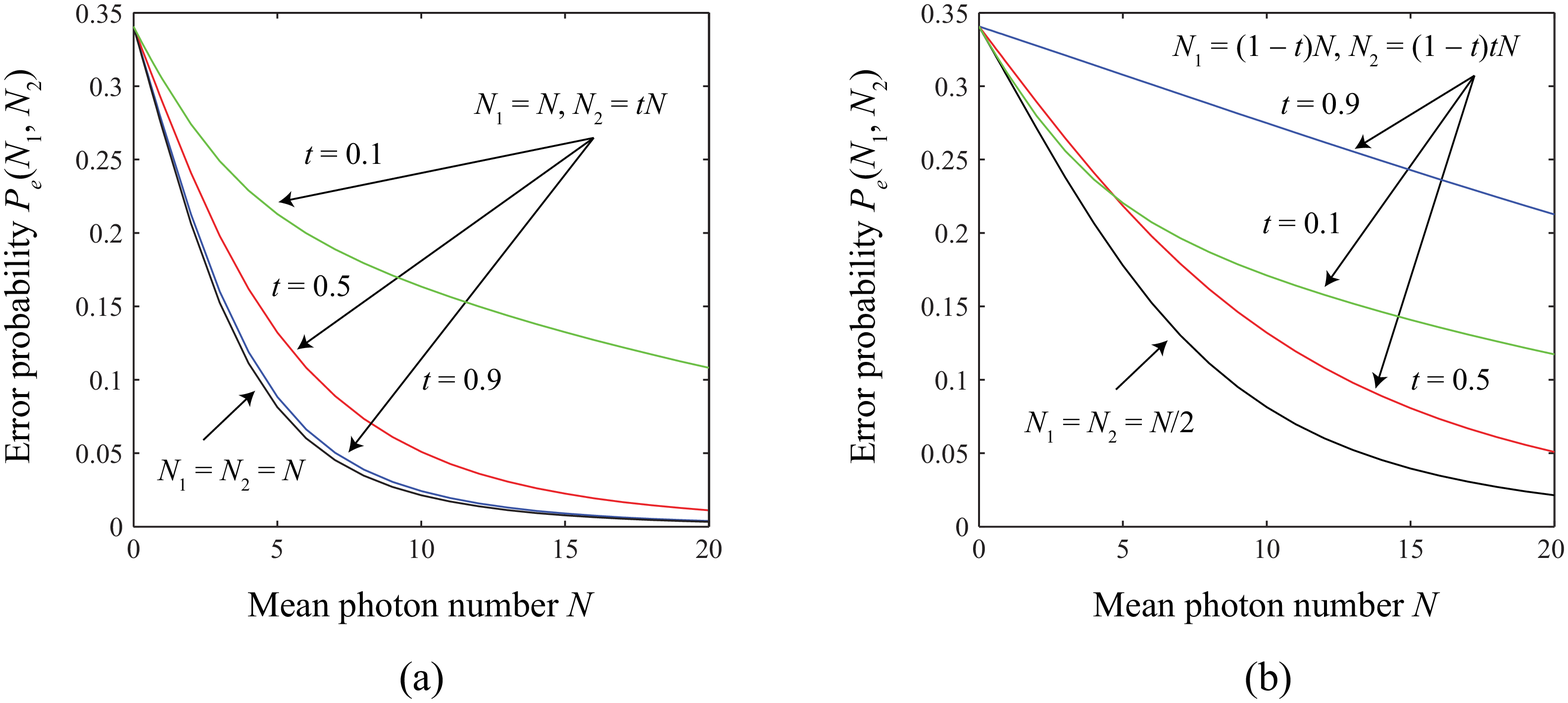}
    \caption{Plots of (a) the IR and (b) the PNS error probabilities $P_e(N_1, N_2)$ of Eve as functions of the mean number of photons $N$ (see the text for detail).}
\label{fig:Figure2_PNS}
\end{figure*}

For the PNS attack, Eve's best strategy without causing errors to Bob's received bits will be to take $N_1 = N_2 \approx N/2$ if Bob did not monitor the photon statistics. Nevertheless, we require that Bob monitors the number of incoming photons so that Eve cannot probe Alice and his devices with very bright pulses. For the IR attack, we can consider $N_1 = N_2 \approx N$. This corresponds to the optimal situation for Eve when the channel is assumed to be lossless. For a lossy channel with transmittance $t$, we consider $N_1 = N$ and $N_2 = tN$ for IR and $N_1 = (1-t)N$ and $N_2 = (1-t)tN$ for PNS. Figure~\ref{fig:Figure2_PNS} gives the plots of $P_e$ as a function of the mean photon number $N$. It is seen in Fig.~\ref{fig:Figure2_PNS} that even at the mean photon number $N = 10$, there is considerable error in Eve's estimated values of the true bit values.

As mentioned previously, Alice and Bob need to monitor the number of incoming photons to avoid Eve from injecting a very bright beam to probe their encoding devices. The presence of Eve is revealed if Alice and Bob also check the photon number distribution and detect any loss or change of the distribution. Eve could compensate the photon loss in the channel by injecting photons of arbitrary polarizations or at the angles $\hat{\phi}_i$, as in the IR attack. Nevertheless, this introduces extra error in her determination of $\hat{\theta}_X$ as well as error in Bob's bits. In addition, the IR attack in fact induces errors to the bit values obtained by Bob. The estimation of the rotation angle error is addressed by the authentication process which specifically handles the man-in-the-middle attack in the next section.

%%%%%%%%%%%%%%%%%%%%%%%%%%%%%%%%%%%%%%%%%%%%%%%%%%%%%%%%%%%%%%%%%%%%%%%%%%%%%%%%%%%%%%
\subsection{Authentication}\label{sec:authentication}
%%%%%%%%%%%%%%%%%%%%%%%%%%%%%%%%%%%%%%%%%%%%%%%%%%%%%%%%%%%%%%%%%%%%%%%%%%%%%%%%%%%%%%
\begin{figure}[!b]
    \centering
    \includegraphics[height=4.7cm]{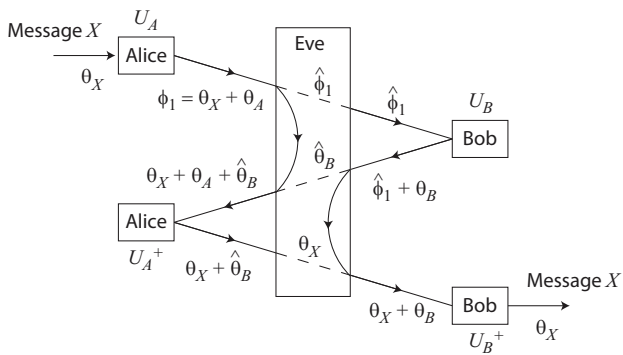}
    \caption{Schematic diagram of the three-stage protocol under the man-in-the-middle (MIM) attack. Eve impersonates Alice and Bob simultaneously. To Alice, Eve impersonates Bob and operates the three-stage protocol by applying her own rotation angle $\hat{\theta}_B$, which could be any angle or the one that she derives from the photons coming from Bob, in the second stage. Likewise she can apply the angle $\hat{\phi}_1$ to the photons sending to Bob in the first stage to minimize the chance of Bob catching her.}
\label{fig:Figure3_MIM}
\end{figure}

The three-stage protocol can be compromised entirely if Eve launches the man-in-middle (MIM) attack as depicted in Fig.~\ref{fig:Figure3_MIM}. Here Eve impersonates Bob to extract the true bit value perfectly. She also impersonates Alice to send the bit angle $\theta_X$ together with the unperturbed angle $\theta_B$ back to Bob, so that Bob receives the bit without error and hence cannot catch Eve. In such an MIM attack, Eve totally separates the quantum communication between Alice and Bob.  Therefore the attack could be revealed if authentications are made by Alice and Bob to guarantee the locks are legitimate, i.e., they are the true users who applied the rotation angles on the pulses they received in the three stages.

In principle, authentication can be performed perfectly if Alice and Bob could retain the photons in steps 2 and 3 of the protocol above until the end of the key exchange, which can be accomplished by using quantum memories~\cite{Lvovsky-etal2009, Sangouard-etal2011} or slow light technologies~\cite{Wu-etal2013}. More practically they need to perform measurements to determine the parameters of the transformations during the key exchange. At the end of the key exchange, they check their measured values against the true values (step 6). It should be noted that we assume Alice and Bob are authenticated for exchanging classical information on a public channel. This will rule out the chance that Eve is also in the middle when Alice and Bob try to compare the measurements.

\begin{figure*}[!t]
    \centering
    \includegraphics[height=8.5cm]{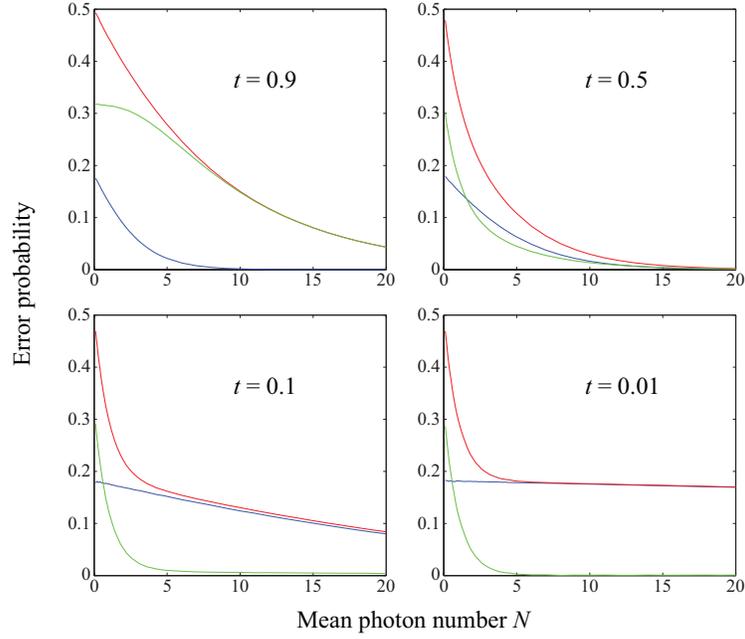}
    \caption{Bob's error probabilities in the estimation of $\theta_X$ for the normal three-stage operation ($P_e^\text{Auth, norm}$, blue lines) and under the MIM attack ($P_e^\text{Auth, MIM}$, red lines) at different values of the channel transmittance $t$. The green lines denote the differences between the two error probabilities.}
\label{fig:Figure4_Authentication}
\end{figure*}

We consider that the transmittance of the quantum channel is $t$. If Alice sends pulses with a mean photon number of $N$, Bob expects to receive pulses with mean photon number $tN$ in the first stage and $t^3 N$ in the third stage, and Alice expects to receive pulses with mean photon number $t^2 N$ in the second stage. Therefore, for the MIM attack, Eve can extract a mean photon number of $(1-t^2)N$ to obtain the estimate $\hat{\phi}_1$ and a mean photon number of $t(1-t^2)N$ to obtain the estimate $\hat{\theta}_B$. Eve then use these two angles to impersonate Alice and Bob simultaneously. If Bob uses the pulse for authentication instead of the normal three-stage, the angle $\tilde{\phi}_1$ he measures conditioned on $\phi_1$ will have a distribution given by $P_1(\tilde{\phi}_1 | \hat{\phi}_1, tN) P_1(\hat{\phi}_1 | \phi_1, (1-t^2)N)$. Here $\phi_1$ is announced to Bob by Alice at the end of the protocol. Using this angle, Bob can guess $\theta_X$ with an error probability of
\begin{eqnarray}
    P_e^\text{Auth, MIM}(t, N)
    \hspace{-1mm}
    &=&
    \hspace{-1mm}
    \int_{\left|\phi_1 - \tilde{\phi}_1\right| > \pi/4}
    d\phi_1 \, d\tilde{\phi}_1 \, d\phi_2 \, d\tilde{\phi}_2 \
\nonumber\\
&&\hspace{-20mm}
    P_1(\tilde{\phi}_1 | \hat{\phi}_1, tN) P_1(\hat{\phi}_1 | \phi_1, (1-t^2)N)
    P(\phi_1)
    .
\label{eq:Pe_AuthMIM}
\end{eqnarray}
On the other hand, in the normal operation when the MIM attack is not present, Bob's error probability is instead given by
\begin{eqnarray}
&&\hspace{-8mm}
    P_e^\text{Auth, norm}(t, N)
\nonumber\\
    \hspace{-1mm}
    &=&
    \hspace{-1mm}
    \int_{\left|\phi_1 - \tilde{\phi}_1\right| > \pi/4}
    d\phi_1 \, d\tilde{\phi}_1 \
    P_1(\tilde{\phi}_1 | \phi_1, tN)
    P(\phi_1)
    .
\label{eq:Pe_AuthNormal}
\end{eqnarray}
It is remarked that Eqs.~(\ref{eq:Pe_AuthMIM}) and (\ref{eq:Pe_AuthNormal}) manifest that fact that, like Eve, Bob and Alice cannot estimate the polarization angles with certainty in the middle of the three-stage protocol because the photons are not in orthogonal states.

Numerical simulations were performed using the measurement scheme described in the last section. Figure~\ref{fig:Figure4_Authentication} shows the two error probabilities as functions of the mean photon number $N$ for different values of the transmittance $t$. In addition, $P_e^\text{Auth, norm}$ is found analytically to be
\begin{eqnarray}
&&\hspace{-8mm}
    P_e^\text{Auth, norm}(t, N)
\nonumber\\
    \hspace{-1mm}
    &\approx&
    \hspace{-1mm}
    \frac{2}{\pi\left(1-e^{-N}\right)}
    \Bigg[
    \int_{\pi/4}^{\pi/2}
    \sum_{n_H = 1}^\infty P_1(n_H, 0| \phi_1, tN) d\phi_1
\nonumber\\
&&
    +
    \int_{0}^{\pi/4}
    \sum_{n_V = 1}^\infty P_1(0, n_V| \phi_1, tN) d\phi_1
    \Bigg]
%\nonumber\\
%    \hspace{-1mm}
%    &=&
%    \hspace{-1mm}
%    \frac{4}{\pi} \int_0^{\pi/4} \frac{1-e^{tN\sin^2\phi_1}}{1-e^{tN}} d\phi_1
\nonumber\\
    \hspace{-1mm}
    &=&
    \hspace{-1mm}
    \frac{e^{tN/2}\left[I_0\left(\frac{tN}{2}\right) - L_0\left(\frac{tN}{2}\right)\right] - 1}
    {e^{tN} - 1}
    ,
\end{eqnarray}
where $I_n(x)$ is the modified Bessel function of the first kind and $L_n(x)$ is the modified Struve function.

It is noted in Fig.~\ref{fig:Figure4_Authentication} that at small $N$ the error probabilities tend to the constant values $P_e^\text{Auth, norm} \rightarrow 2^{-1} - \pi^{-1}$ and $P_e^\text{Auth, MIM} \rightarrow 0.5$ whereas both probabilities tend to zero at large $N$. When the transmittance decreases, the two error probabilities converge to each other at a smaller $N$. In addition, the difference $P_e^\text{Auth, MIM} - P_e^\text{Auth, norm}$ approaches an asymptotic form when $t \rightarrow 0$, which is non-negligibly greater than zero for $N < 4$.

%%%%%%%%%%%%%%%%%%%%%%%%%%%%%%%%%%%%%%%%%%%%%%%%%%%%%%%%%%%%%%%%%%%%%%%%%%%%%%%%%%%%%%
\subsection{Secure Key Rate and Rate Efficiency}\label{sec:key_rate}
%%%%%%%%%%%%%%%%%%%%%%%%%%%%%%%%%%%%%%%%%%%%%%%%%%%%%%%%%%%%%%%%%%%%%%%%%%%%%%%%%%%%%%
With error correction and privacy amplification, the expression for the secret key rate extractable using one-way classical postprocessing is~\cite{Scarani-etal2009}
\begin{equation}
    K = R \left[ I(A:B) - \min\left(I_{EA}, I_{EB}\right)\right] ,
\end{equation}
where $R$ is the raw key rate, $I(A:B)$ is the mutual information between Alice and Bob, and $I_{EA}$ and $I_{EB}$ are Eve's information about the raw key of Alice and Bob respectively. We consider the case when $H(A) = H(B) = 1$ and $H(A|B) = H(B|A) = h(Q)$, where $h(Q)$ is the binary entropy function and $Q$ is the quantum bit error rate (QBER). For the three-stage protocol, the raw key rate is given by the total bit that Bob measured minus the bits for authentication.

Assuming the error correction is carried out perfectly and using a very conservative estimate for the PNS/IR attack mentioned in Section~\ref{sec:IR_PNS} with mutual information $I(E:A) = 1 - h\left(P_e(N, tN)\right)$, the security key rate then becomes
\begin{equation}
    K = R \left[(1-f)h\left(P_e(N, tN)\right) - h(Q)\right] ,
\end{equation}
where $f$ is the fraction of the MIM attacks launched by Eve, which is estimated by the ratio of the measured authentication error probability difference and the expected measured authentication error probability difference, i.e.,
\begin{equation}
    f =
    \frac{P_e^\text{Auth, measured}(t, N) - P_e^\text{Auth, norm}(tN)}
    {P_e^\text{Auth, MIM}(t, N) - P_e^\text{Auth, norm}(tN)} .
\end{equation}
The threshold for the QBER is then determined by the condition $K > 0$ for some given $f < 1$ and $N$.

A potentially significant drawback of the three-stage protocol compared to other QKD protocols is that it requires multiple quantum communications between Alice and Bob, effectively increases the photon loss of the channel. On the other hand, the multiple-photon resilient nature of the protocol allows a larger mean photon number to start with. As an estimate, we consider the ratio of the raw bit rates between the three stage protocol and the weak-coherent state BB84 with mean photon number 0.5. The ratio is given by
\begin{equation}
    E = \frac{1 - e^{N t(2l)}}{1 - e^{-0.5 t(l)}} ,
\end{equation}
where the transmittance $t$ is given as a function of distance by $t(l) = 10^{-\alpha l/10}$. For optical fiber at the wavelength 1550 nm, $\alpha = 0.2~\text{dB/km}$. The three-stage protocol will have advantage over a one-stage protocol if $E > 1$. This gives
\begin{equation}
    l \le \left(\frac{5}{\alpha} \log_{10} \frac{N}{0.5}\right) \text{km} .
\end{equation}

It is emphasized that the choice of $N$ depends on the security level described in the previous sections. From the previous analysis, it is seen that the MIM attack puts a more stringent condition to $N$ than the PNS/IR attack. It is estimated that for up to a distance of 20 km with $N = 3$, the three-stage protocol could be advantageous over the decoy state BB84.

%%%%%%%%%%%%%%%%%%%%%%%%%%%%%%%%%%%%%%%%%%%%%%%%%%%%%%%%%%%%%%%%%%%%%%%%%%%%%%%%%%%%%%
\subsection{Amplification Attack}\label{sec:amplification_attack}
%%%%%%%%%%%%%%%%%%%%%%%%%%%%%%%%%%%%%%%%%%%%%%%%%%%%%%%%%%%%%%%%%%%%%%%%%%%%%%%%%%%%%%
So far we have only focused on the situation where Eve makes direct measurement using the photons that she siphons off from the quantum channel. Generally she can do more with her photons. An important class of attacks is by amplifying the quantum states that she extracts from the channel. This kind of attack is linked to the foundation of the three-stage protocol, that is whether she can find out the angles $\theta_A$ and $\theta_B$, which are open to her eavesdropping, with high precision. In fact, the purpose of using finite number of photons in the channel is to limit Eve's precision of measurement.

It is well known that the amplification of a quantum state must also accompany with the amplification of the noise~\cite{Caves1982}. For the implementation with coherent states discussed in this paper, Eve does not gain anything by amplifying the signal. Even with the use of squeezed states, Fock states or entangled states to resend pulses to Bob and Alice, the intensity check by Alice and Bob will introduce vacuum noise to Eve's probes, and Eve's information gain may only be modest.

On the other hand, it is recently shown that noiseless amplification of a quantum state is possible if a perfect guarantee of success is not required, unlike the usual deterministic linear amplification mentioned above~\cite{Ralph-Lund2009, Pandey-etal2013}. Experiments of amplifying coherent states noiselessly have already been demonstrated~\cite{Xiang-etal2010,Zavatta-etal2011}. This apparently imposes a significant drawback to the three-stage protocol. Nevertheless, it should be noted that the probabilistic nature of the amplification means that Eve's bit rate will further decrease. More importantly, the implementations of the amplification of coherent states operate with high fidelity only when the mean photon number after the gain is around unity~\cite{Xiang-etal2010,Zavatta-etal2011}. The amplification attack works well essentially for very weak coherent states but not for the regime of $N > 1$ that we consider in our protocol. The distortion of the quantum states at larger $N$ introduces noise to the determination of the polarization. Further work is needed to quantify the effects of the amplification attack to the security of the protocol.

Another issue related to the amplification attack on the three-stage protocol is that, in the actual implementation, the polarization rotations $\theta_A$ and $\theta_B$ nevertheless have to be confined to a finite set because of the noise and stability of the experimental setup. Such limitation may open up the unambiguous state discrimination (USD) attack~\cite{vanEnk2002, Becerra-etal2013}. Fortunately, the polarization rotations are local information that is secret to Alice and Bob independently; they can change their sets of the rotations frequently without disclosing their actions. This results in an extremely large set of the polarization rotations and effectively mitigates the threat of the USD attack, which requires that the number of photons needed must be greater than or equal to the number of polarization states in the middle of the three-stage protocol.

%%%%%%%%%%%%%%%%%%%%%%%%%%%%%%%%%%%%%%%%%%%%%%%%%%%%%%%%%%%%%%%%%%%%%%%%%%%%%%%%%%%%%%
\section{Conclusion}\label{sec:conclusion}
%%%%%%%%%%%%%%%%%%%%%%%%%%%%%%%%%%%%%%%%%%%%%%%%%%%%%%%%%%%%%%%%%%%%%%%%%%%%%%%%%%%%%%
This paper gave a detailed security analysis to a new form of quantum key distribution protocol, the three-stage multi-photon quantum cryptography system, using coherent states to encode qubits. In particular, we showed that the three-stage protocol is resilient to the photon number splitting attack, the intercept-resend attack, and the man-in-the-middle attack with the error probabilities calculated as functions of the mean number of photons in the channel. We have obtained the secure key rate in terms of the error probabilities under the attacks considered. Importantly, we found that the mean photon number of the coherent states can practically be larger than 1, in contrast to most current QKD protocols in which weak coherent pulses (mean photon number $\sim 0.1$ for BB84 to 0.6 for decoy-BB84) are considered. The multi-photon multi-stage scheme presented does not require pre-sharing of a key between the legitimate users like the Y00 protocol. Hence it can be used to complement such multi-photon quantum communication protocols.

We have also discussed the amplification and unambiguous state discrimination attacks and argued that such attacks do not impose significant threat to our protocol. Further study on this issue will be pursued in future to quantify the actual effect of the amplification attack using the analytic tool for the optimal coherent state amplifier~\cite{Chiribella-Xie2013}.

%%%%%%%%%%%%%%%%%%%%%%%%%%%%%%%%%%%%%%%%%%%%%%%%%%%%%%%%%%%%%%%%%%%%%%%%%%%%%%%%%%%%%%%%%%%
\begin{acknowledgments}
This research is supported in part by the National Science Foundation (NSF) under Grants 1117148, 1117179, and 1117068.
\end{acknowledgments}

%%%%%%%%%%%%%%%%%%%%%%%%%%%%%%%%%%%%%%%%%%%%%%%%%%%%%%%%%%%%%%%%%%%%%%%%%%%%%%%%%%%%%%%%%%%

\end{document}